\documentclass[12pt]{iopart}
\usepackage{graphicx,bm}
\usepackage{iopams}  
\usepackage{bm}

\newcommand{\be}{\begin{eqnarray}}
\newcommand{\ee}{\end{eqnarray}}
\newcommand{\ba}{\begin{array}}
\newcommand{\ea}{\end{array}}
\newcommand{\no}{\nonumber}
\newcommand{\lb}{\left(}
\newcommand{\rb}{\right)}
\newcommand{\lbb}{\left\{}
\newcommand{\rbb}{\right\}}
\newcommand{\lsb}{ \left[ }
\newcommand{\rsb}{ \right] }

\newcommand{\mR}{\mathbb{R}}
\newcommand{\mN}{\mathbb{N}}

\begin{document}

\title[Statistical mechanical analysis of a hierarchical random code ensemble]
{Statistical mechanical analysis of a hierarchical random code ensemble
in signal processing}

\author{Tomoyuki Obuchi$^1$, Kazutaka Takahashi$^2$ and Koujin Takeda$^3$}

\address{
$^1$ Department of Earth and Space Science, Faculty of Science, \\
 Osaka University, Toyonaka 560-0043, Japan

$^2$ Department of Physics, 
 Tokyo Institute of Technology, Tokyo 152-8551, Japan

$^3$ Department of Computational Intelligence and Systems Science, \\
 Tokyo Institute of Technology, Yokohama 226-8502, Japan
}
\begin{abstract}
 We study a random code ensemble with a hierarchical structure, 
 which is closely related to the generalized random energy model 
 with discrete energy values. 
 Based on this correspondence, we analyze the hierarchical
 random code ensemble 
 by using the replica method in two situations: 
 lossy data compression and channel coding. 
 For both the situations, 
 the exponents of large deviation analysis
 characterizing the performance of the ensemble, 
 the distortion rate of lossy data compression
 and the error exponent of channel coding in Gallager's formalism, 
 are accessible by a generating function 
 of the generalized random energy model.
 We discuss that the transitions of those exponents 
 observed in the preceding work can be interpreted as
 phase transitions with respect to the replica number.
 We also show that the replica symmetry breaking plays an essential role 
 in these transitions.
\end{abstract}

\maketitle

\section{Introduction}

 Signal processing is one of the main topics
 in information science and gaining much more significance
 in modern society. 
 In this connection, 
 statistical mechanical approaches to signal processing
 have been investigated for decades, which have provided
 various novel viewpoints to information theory~\cite{Nishimoribook,MM}.

 Among various models in information theory, 
 the random code ensemble is known as a fundamental model.
 This ensemble was introduced by
 Shannon~\cite{Shannon1948a,Shannon1948b}
 and found to show the optimal performance in error correction 
 stated in the channel coding theorem investigated by himself.
 After the original study, Gallager~\cite{Gallager1965} 
 enforced its significance through the perfection
 of Shannon's result.
 In the context of statistical mechanics, this ensemble
 can be viewed as a fundamental spin-glass model: 
 in certain limits, this corresponds to
 the random energy model (REM) proposed and rigorously analyzed 
 by Derrida~\cite{Derrida1980,Derrida1981}. 
 This relation was first pointed out by Sourlas~\cite{Sourlas1989}. 
 His work has been recognized as an epoch-making result followed by 
 numerous works such as~\cite{Rujan,Sourlas1994,KabashimaSaad1998} 
 about decoding
 and~\cite{KanterSaad,VicenteSaadKabashima,
 KabashimaMurayamaSaad,MontanariSourlas,Montanari2000} 
 about performance-achieving code.

 As a generalization of the REM,
 the model with a hierarchical structure, 
 termed the generalized random energy model (GREM), was also proposed
 and rigorously solved in~\cite{Derrida1985, DerridaGardner1,
 DerridaGardner2}. 
 The original motivation of the generalization was to clarify 
 the relation of the GREM with the other mean-field spin glass model 
 such as the Sherrington-Kirkpatrick model~\cite{SK}.
 In a recent work, Merhav~\cite{Merhav} proposed
 a random code ensemble with a hierarchical structure
 for performance improvement and argued that
 such a hierarchical ensemble has a similar structure to the GREM. 
 Based on such a similarity,
 he investigated two issues by large deviation analysis: distortion
 in lossy data compression and performance of the Bayesian decoder
 in channel coding through the binary symmetric channel (BSC).
 For lossy data compression, 
 he concluded that for higher performance  
 the parameters describing the hierarchical structure should be tuned 
 to a range where the GREM shows the same thermodynamic behavior 
 as the standard REM. 
 He also discussed that the same tuning of hierarchical parameters
 for optimal performance holds
 in channel coding. 
 However, for a decisive conclusion more detailed investigations
 are desired. As a crucial point, in taking the ensemble average
 for performance evaluation we need to consider quenched average, 
 whereas in his analysis simpler annealed average was adopted
 although he gave some justifications.
 
 Under the circumstances, we reinvestigate the hierarchical random
 code ensemble in a more inclusive way by using the replica method, 
 which enables us to evaluate the performance of the code
 with quenched average.
 In our recent work~\cite{OTT}, 
 we analyzed the GREM by the replica method 
 and found that the multiple-step replica symmetry breaking (RSB)
 appears at low temperatures in the quenched limit. 
 The quenched and the annealed limits are connected with each other
 in a region where a replica number is positive.
 This positive replica region becomes important for the large deviation
 analysis of the random code ensemble. We analyze this region in detail and 
 see that the similar RSB transitions again appear.
 They play a crucial role for the transitions
 of the distortion rate 
 and Gallager's error exponent~\cite{Gallager1965,Gallagerbook}, 
 which directly concerns the performance of the random code. 

 The actual analysis is performed on a generalized discrete random
 energy model (GDREM). This model, where possible values of random
 energy are discrete unlike the original REM,
 can be seen as a generalization of the discrete REM
 in~\cite{OgK1,OgK2,OgK3}. We apply the replica analysis to the GDREM
 and obtain the phase diagram for the region of a non-negative
 replica number.  The GDREM is directly mapped to the hierarchical
 random code ensemble. This mapping enables us to readily interpret
 the properties of the GDREM in the context of the random code.
 Phase transitions involving the higher step RSB 
 found in the GDREM are directly connected to those in the distortion rate
 and Gallager's error exponent. We emphasize that the transitions
 in the region of a positive replica number are not merely theoretical
 matters in the replica analysis, but also have a practical significance
 in information theory. The physical interpretations of behaviors of
 the distortion rate and Gallager's error exponent constitute a part of
 main results in this paper.

 This paper is organized as follows. 
 In section~\ref{sec:GDREM}, we introduce the GDREM and 
 analyze the phase diagram using the replica method.
 We show that many phases coexist on the diagram of 
 temperature versus the replica number.  
 In section~\ref{sec:datacomp}, we briefly review 
 the discussion of distortion in lossy data compression, 
 and compare the result from our replica analysis of the GDREM 
 with~\cite{Merhav}. 
 As shown there, the replica analysis enables us
 to investigate the distortion rate quite readily. The result indicates
 that the higher step RSB degrades the performance of
 a general hierarchical code.
 Error correction by the Bayesian decoder is studied
 in section~\ref{sec:channelcoding},
 where Gallager's error exponent is rederived from our result.
 We show that two-parameter optimization probably becomes significant 
 when correlations between codewords exist.
 We also point out that the concentration of the Gibbs measure can be strongly 
 related to the performance analysis of the Bayesian decoder. 
 The last section is devoted to the conclusion.

\section{The GDREM}
\label{sec:GDREM}
 In this section we introduce and analyze the GDREM.
 The REM~\cite{Derrida1980,Derrida1981}
 is one of the fundamental models in spin glasses,
 and in its definition the energy of respective state is taken as random.
 Derrida and Gardner generalized the REM,
 termed the GREM, in their subsequent works~\cite{Derrida1985,
 DerridaGardner1,DerridaGardner2} by incorporating the hierarchical
 structure in the random energy.
 In the original work of the REM or the GREM,
 the probability distribution of the energy is Gaussian, whereas
 the GDREM dealt with here is the model of discrete random energy.
 In the following we study the GDREM with the binomial distribution 
 of hierarchical random energy.

\subsection{Random variable representation}
\label{sec:energyrep}

\begin{figure}[htb]
\begin{center}
\includegraphics[width=0.75\columnwidth]{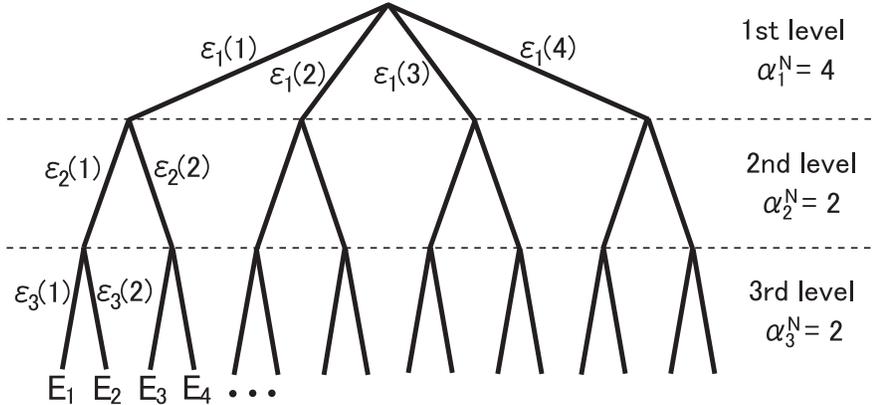}
\caption{Schematic picture of hierarchical random energy.
 Here the case of $K=3, N=4$ and $\{\alpha_1^N, \alpha_2^N \alpha_3^N\}
 = \{ 4,2,2 \}$ is depicted. From the root to a leaf
 (corresponding to a state $i$) of the tree,
 we sum up $\epsilon_{\nu}(j)$, which becomes the energy of
 the $i$th state $E_i$.}
\label{fig:hierarchicalenergy}
\end{center}
\end{figure}

 First we give the definition of the GDREM.
 We follow the notation for the GREM in our paper~\cite{OTT}.
 Prepare $K$ hierarchical levels, and
 for the $\nu$th level $(1\le \nu \le K)$ random variables
 $\epsilon_{\nu}(1), 
 \epsilon_{\nu}(2), \ldots, \epsilon_{\nu}(M_{\nu})$
 are assigned. These random variables $\{\epsilon_{\nu}\}$ become
 the energy components of the $\nu$th hierarchy.
 The number of independent random variables for the $\nu$th level,
 $M_{\nu}$, can be factored as
\be
 \label{defM}
 M_{\nu} = (\alpha_1 \cdots \alpha_{\nu})^N,
\ee
 where $\alpha_{\nu}^{N}$ is an integer satisfying
 $1<\alpha_{\nu}^{N}<2^N$ and denotes the number of independent
 random variables $\{\epsilon_{\nu} \}$ belonging to a state in the
 $(\nu-1)$st level (see figure~\ref{fig:hierarchicalenergy}).
 For the deepest level $\nu=K$, $M_{K}=(\alpha_1 \cdots
 \alpha_{K})^N=2^N$ must be held. 

 From the random variables, we introduce $2^N$ new variables
 $\{E_{i}\}$, which represent the energy of the system and are defined as 
\be
\label{energyrep}
 E_i  = \sum_{\nu=1}^{K} \epsilon_{\nu} 
 (\lfloor(i-1)M_{\nu}/2^N\rfloor+1)
  = \sum_{\nu=1}^{K} \epsilon_{\nu}^{(i)}, 
\ee
 where $i=1,\ldots, 2^N$ and $\lfloor x \rfloor$ denotes 
 the floor function indicating the largest integer not exceeding $x$.
 This structure is depicted in figure~\ref{fig:hierarchicalenergy}.

 Then, the partition function is defined by
\be
\label{eq:partition1}
 Z(\beta) = \sum_{i=1}^{2^N} e^{-\beta E_i} = \sum_{i=1}^{2^N}
 \exp \left( -\beta \sum_{\nu=1}^{K} \epsilon_{\nu}^{(i)} \right), 
\ee
 with $\beta=1/T$ being the inverse temperature.

 The properties of this model are determined by the distribution
 of the random variables $\{ \epsilon_{\nu} \}$. 
 Here we choose the binomial distribution to see the connection
 with the random code ensemble. 
 For the $\nu$th level, the distribution is characterized 
 by a parameter $L_\nu$. The specific form is 
\be
\label{binomialdist}
 P_{\nu}(\epsilon_{\nu}) 
 = \sum_{l=0}^{L_\nu}\frac{L_{\nu}!}{l! (L_{\nu} - l)!} 
 \left( \frac{1}{2} \right)^{L_{\nu}}
 \delta\left(\epsilon_{\nu},l - \frac{L_\nu}{2} \right),
\ee
 where $\delta(x,y)$ is the Kronecker delta function.
 The number of possible values of $\epsilon_{\nu}$ is $L_{\nu}+1$, 
 namely, $\epsilon_\nu$ can take any of  
 $-L_{\nu}/2, -L_{\nu}/2+1, \ldots, L_{\nu}/2-1, L_{\nu}/2$.
 For later convenience, we define the parameters 
 $a_{\nu}=L_{\nu}/N$ and $a=\sum_{\nu=1}^{K}a_{\nu}$.
 In contrast to the case of the Gaussian REM,
 the value of the parameter $a$ is significant.
 The RSB occurs only for $a>1$ as discussed in~\cite{OgK1},
 in which case we study in the following.

\subsection{Bit representation}
\label{sec:bitrep}

 Here we give another definition of the GDREM by using bit variables
 to deal with the hierarchical random code ensemble.

 Prepare $aN$ bits
 taking the value 0 or 1, and divide them into $K$ blocks as
 $aN=\sum_{\nu=1}^{K} L_{\nu}$.
 For the $\nu$th block composed of $L_{\nu}$ bits, 
 we randomly choose $M_{\nu}$ bit configurations from possible
 $2^{L_{\nu}}$ ones, denoted by ${\bm z}_{\nu}(1), {\bm z}_{\nu}(2), 
 \ldots, {\bm z}_{\nu}(M_{\nu})$ 
 where each ${\bm z}_{\nu}$ has $L_\nu$ components.
 The $i$th configuration ${\bm x}_i$ is expressed by arraying 
 the respective configuration of each block as
\be
 \bm x_i = \left\{
 {\bm z}_1(\lfloor (i-1)M_{1}/2^N \rfloor +1),
 \ldots, 
 {\bm z}_K(\lfloor (i-1)M_{K}/2^N \rfloor +1)
 \right\},
\ee
 namely ${\bm x}_i$ is composed of $aN$ elements. This procedure 
 constructs $2^N$ bit configurations from $2^{aN}$ possible ones. 
 The resultant set of chosen configurations, 
 which is denoted as $\cal{C}$ hereafter, has a hierarchy with $K$
 levels as the random variable representation.

 After construction of states, 
 we define the Hamming distance which counts 
 the number of different bits 
 between bit sequences $\bm x_i$ and ${\bm y}$ as
\be
 d_{\rm H}(\bm x_i, {\bm y}) 
 = \sum_{\nu=1}^{K} 
 \sum_{l=1}^{L_{\nu}}
 \delta (
 {\bm z}_{\nu}^{(l)}(\lfloor (i-1)M_{\nu}/2^N \rfloor +1), 
 {\bm y}_\nu^{(l)}).
\ee
 Here, $\bm z_\nu^{(l)}$ 
 is the $l$th component of the bit configuration ${\bm z}_\nu$, 
 and ${\bm y}$ is a reference bit sequence 
 which may be chosen as the simplest one 
 such as the all-zero sequence ${\bm 0}$.
 Using the Hamming distance, 
 we define the energy of the $i$th state as 
 $E_i = d_{\rm H}(\bm x_i ,\bm 0) - aN/2 $, which leads to  
  the partition function of this system as
\be
\label{eq:partition}
 Z(\beta)=\sum_{{\bm x}\in \cal{C}}
 \exp\lbb -\beta 
 \lb d_{\rm H}({\bm x},{\bm 0})-\frac{aN}{2} \rb
 \rbb.
\ee
 The ensemble of bit sequences given here is nothing but
 the hierarchical random code ensemble 
 introduced in~\cite{Merhav}.
 The energy has the same hierarchical structure 
 as in (\ref{energyrep}) and
 the energy of each block is drawn from
 the binomial distribution in (\ref{binomialdist}), which means 
 that the representation (\ref{eq:partition}) is equivalent
 to (\ref{eq:partition1}).
 As we see later, this expression is convenient for 
 the discussion of signal processing.

\subsection{Replica analysis}
\label{sec:replica}

 We analyze the GDREM by the replica method. 
 As is well known, the replica method is a tool for taking ensemble
 average of logarithm or arbitrary power of the partition function.
 This method is thus quite suitable for the performance evaluation
 of the hierarchical random code ensemble, because the ensemble average
 of the arbitrary power of the partition function is totally desired,
 as we see in the following sections.
 The scheme is the same as demonstrated in~\cite{OTT}.
 The difference is only in the probability distribution function of energy. 
 We briefly sketch the main result here.

 Let us evaluate replicated partition function $Z^n$ of the GDREM.
 When $n$ is a natural number, $Z^n$ can be written as
\be
 Z^{n}(\beta) &=& 
 \sum_{i_1=1}^{2^N} \cdots \sum_{i_n=1}^{2^N}
 \exp \left( -\beta \sum_{\nu=1}^{K} (\epsilon_{\nu}^{(i_1)}
 +\epsilon_{\nu}^{(i_2)}+\cdots+\epsilon_{\nu}^{(i_n)})\right)
 \no \\
 &=& \sum_{i_1=1}^{2^N} \cdots \sum_{i_n=1}^{2^N}
 \exp \left( -\beta \sum_{\nu=1}^{K} \sum_{j=1}^{M_\nu}
 n_{\nu} ( j, \{ i_a \}) \epsilon_{\nu} (j)  \right), 
\ee
 where
\be
 n_{\nu} ( j, \{ i_a \} ) = \sum_{a=1}^{n} I_{\nu} (j,i_a),
\ee
 and $I$ is the indicator function
\be
 I_{\nu} (j, i_a) = \left\{
 \ba{cc}
 1 & {\rm for} \  j=  \lfloor (i_a-1) M_{\nu}/2^N \rfloor +1 \\
 0 & {\rm otherwise.}
 \ea\right.
\ee
 The ensemble average yields
\be
\label{eq:averagepartition}
 [Z^{n}] &=& \sum_{i_1=1}^{2^N} \cdots \sum_{i_n=1}^{2^N}
 \exp \left( N
 \sum_{\nu=1}^{K} \sum_{j=1}^{M_{\nu}} a_{\nu}
 \ln \cosh \frac{\beta n_{\nu}(j,\{i_a\})}{2} \right)
 \no\\
 &=& \sum_{\{ n_{\nu} \}} 
 \exp \left( S(\{n_{\nu}\})+N
 \sum_{\nu=1}^{K} \sum_{j=1}^{M_{\nu}} a_{\nu}
 \ln \cosh \frac{\beta n_{\nu}(j,\{i_a\})}{2} \right),
\ee
 where $[\ ]$ means ensemble average and $S(\{n_{\nu}\})$ is 
 the entropy function defined as the logarithm of the number of 
 configurations giving $\{n_{\nu}\}$.
 In deriving (\ref{eq:averagepartition}), we should take care that
 the distribution of the energy is binomial,
 which is only the difference from our preceding work~\cite{OTT}.
 In the thermodynamic limit $N\to \infty$, 
 we need to calculate the saddle-point contribution of $[Z^n]$. 
 A generating function 
 $\phi(\beta, n) \equiv \lim_{N \rightarrow \infty}\ln [Z^{n}]/N$ 
 is convenient for this purpose, 
 and is also significant for signal processing as seen later. 
 In the rest of this section, we focus on calculating 
 this generating function $\phi(\beta,n)$. 
 Hereafter we restrict ourselves to the cases of 
 $K=1, 2$ for simplicity.
 
 Practically, we need $\phi(\beta,n)$ for general $n\in \mR$,
 even though the expression (\ref{eq:averagepartition})
 is valid only for $n \in \mN$.
 To bridge the gap, we utilize the replica method for analytic
 continuation from the natural to real number
 with the Parisi ansatz~\cite{Parisi1980a,Parisi1980b,Parisi1980c}.
 For readers not familiar with these procedures,
 we refer to~\cite{Nishimoribook}. Here we demonstrate a part of
 calculations for the case $K=2$.
 
\begin{figure}[htb]
\begin{center}
\includegraphics[width=0.80\columnwidth]{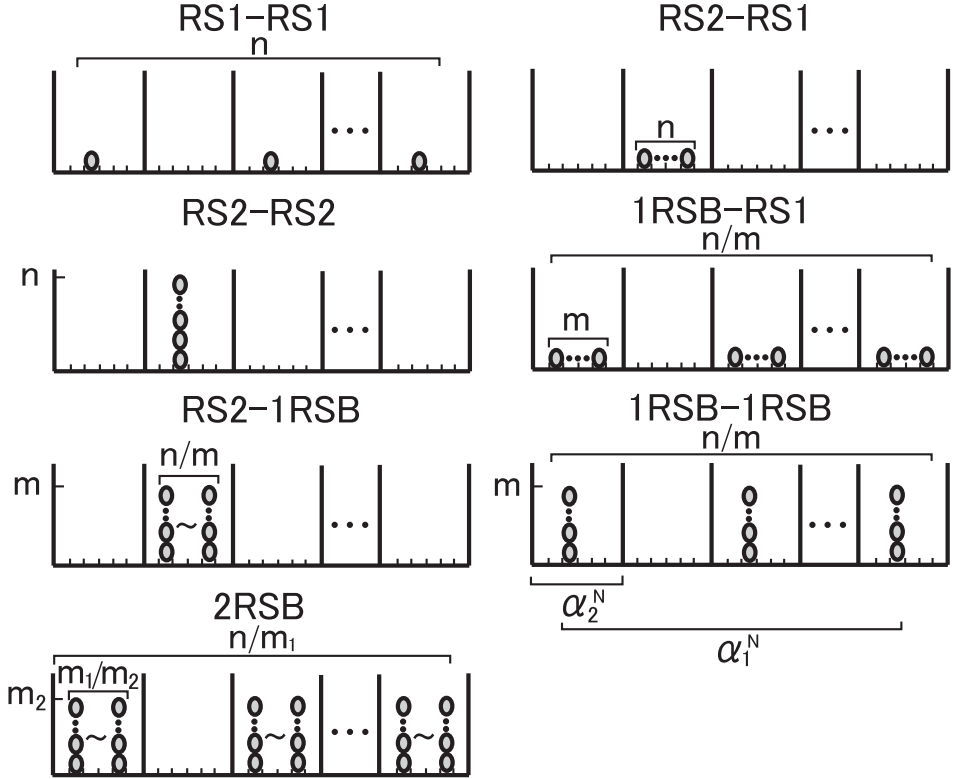}
\caption{Graphical representation of the possible saddle-point 
 solutions at $K=2$. 
 The horizontal and vertical axes represent the index of 
 configurations and the number of ``balls'', respectively.
 All configurations are divided into $\alpha_1^N$-groups 
 including $\alpha_2^N$-configurations. 
 }
\label{fig:saddlepoint}
\end{center}
\end{figure}

 According to the standard prescription using the Parisi ansatz,
 it is sufficient for the current case to consider 
 the replica symmetric (RS) and the one-step RSB (1RSB) solutions 
 in each hierarchy~\cite{OTT,OgK1}.
 If the 1RSB occurs in both the hierarchies with different block sizes, 
 it can be interpreted as the two-step RSB (2RSB).  
 Each solution can be graphically expressed  by how $n$ ``balls'' 
 are partitioned into $2^N$ ``boxes'' (figure \ref{fig:saddlepoint}). 

 For each hierarchy, there are two RS solutions:  
 the RS solutions of the first and the second sorts
 (RS1 and RS2, respectively). 
 For the RS1 solution all $n$ balls are distributed to 
 different states in the hierarchy, 
 while for the RS2 solution all $n$ balls 
 are in the same state in the hierarchy. 
 For example, the RS2-RS1 solution corresponds to 
 the solution being RS2 in the first hierarchy 
 and RS1 in the second one.
 The entropy of this solution is calculated as 
\be 
 S(\{n_{\nu} \})=\ln \{ \alpha_1^{N}\alpha_2^{N}(\alpha_2^{N}-1)\cdots 
 (\alpha_2^{N}-(n-1)) \} \sim N(\ln \alpha_1 + n \ln \alpha_2 ),
 \no\\
\ee
 and the energetic term becomes 
\be 
 \sum_{\nu=1}^{K} \sum_{j=1}^{M_{\nu}} a_{\nu}
 \ln \cosh \frac{\beta n_{\nu}(j,\{i_a\})}{2}
 =a_1 \ln \cosh \frac{\beta n}{2}+n a_2 \ln \cosh \frac{\beta}{2}.
\ee
 These yield the generating function $\phi(\beta,n)$ as
\be
 \phi(\beta,n)=
 \ln \alpha_1 + a_1 \ln \cosh \frac{\beta n}{2}+
 n \left( \ln \alpha_2 + a_2 \ln \cosh \frac{\beta}{2} 
 \right).
\ee
 The other solutions are similarly evaluated; 
 therefore, we skip the derivation.  
 For the RSB solutions, there exist additional parameters 
 (such as $m$ and $m_{1,2}$). 
 These parameters are chosen to extremize $\phi(\beta,n)$ and 
 the explicit dependence on those parameters vanishes 
 in the final step. 
 The possible solutions are summarized as follows:
\be
\label{eq:phiK2}
 \phi(\beta, n) =  \left\{
 \ba{ll}
 n \left( \ln 2 + a \ln \cosh \frac{\beta}{2} \right)
 & {\rm (RS1-RS1)} \\
 \ln \alpha_1 + a_1 \ln \cosh \frac{\beta n}{2}
 & \\
 \ \ \ \ \ + n \left( \ln \alpha_2 + a_2 \ln \cosh \frac{\beta}{2} \right)
 & {\rm (RS2-RS1)} \\
 { \ln 2 + a \ln \cosh \frac{\beta n}{2}}
 & {\rm (RS2-RS2)} \\
 { \frac{a_1 \beta n}{2} \tanh \frac{\beta_1}{2}
 + n \left( \ln \alpha_2 + a_2 \ln \cosh \frac{\beta}{2} \right)}
 & {\rm (1RSB-RS1)} \\
 { \ln \alpha_1 + a_1 \ln \cosh \frac{\beta n}{2}
 + \frac{a_2 \beta n}{2} \tanh \frac{\beta_2}{2}}
 & {\rm (RS2-1RSB)} \\
  {\frac{a \beta n }{2} \tanh \frac{\beta_{\rm c}}{2} }
 & {\rm (1RSB-1RSB)} \\
 { \frac{\beta n}{2} \left( a_1 \tanh \frac{\beta_1}{2} +
 a_2 \tanh \frac{\beta_2}{2} \right)}
 & {\rm (2RSB)} \\
 \ea \right., \no\\ 
\ee
 where the critical temperature $\beta_{\rm c}$ is defined by 
 the equation
\be
\label{eq:transitionK}
 R + 
 \ln \cosh \frac{\beta_{\rm c}}{2}
 - \frac{\beta_{\rm c}}{2} \tanh \frac{\beta_{\rm c}}{2} =0,
\ee
 with $R\equiv \ln 2/a$.
 Other critical temperatures $\beta_1$ and $\beta_2$ are defined by 
 the same equation (\ref{eq:transitionK}) with substitutions 
 $R=\ln \alpha_1/a_1\equiv R_1$ and $R=\ln \alpha_2/a_2\equiv R_2$, 
 respectively.
 
\begin{figure}
\includegraphics[width=0.50\textwidth]{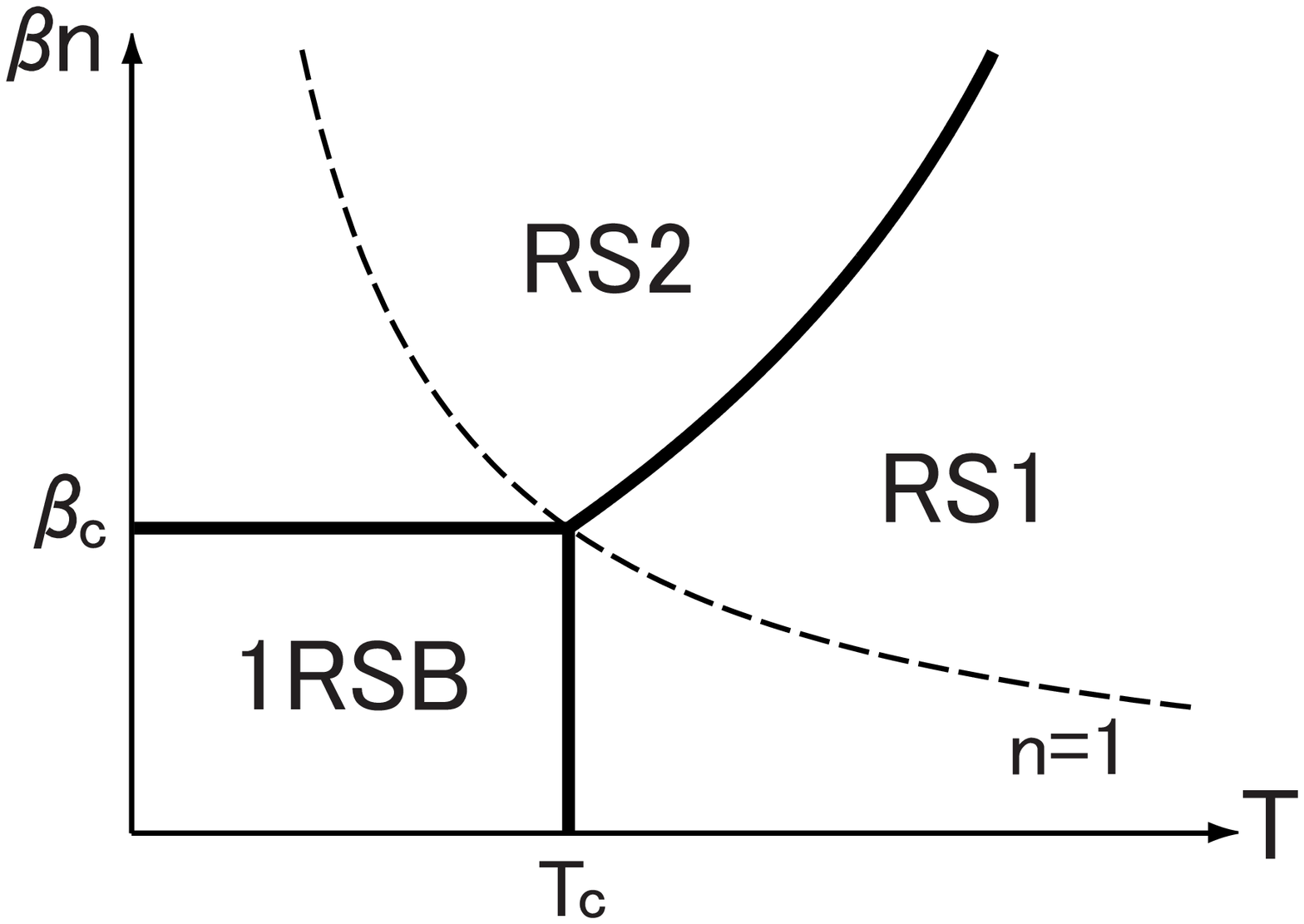}
\includegraphics[width=0.50\textwidth]{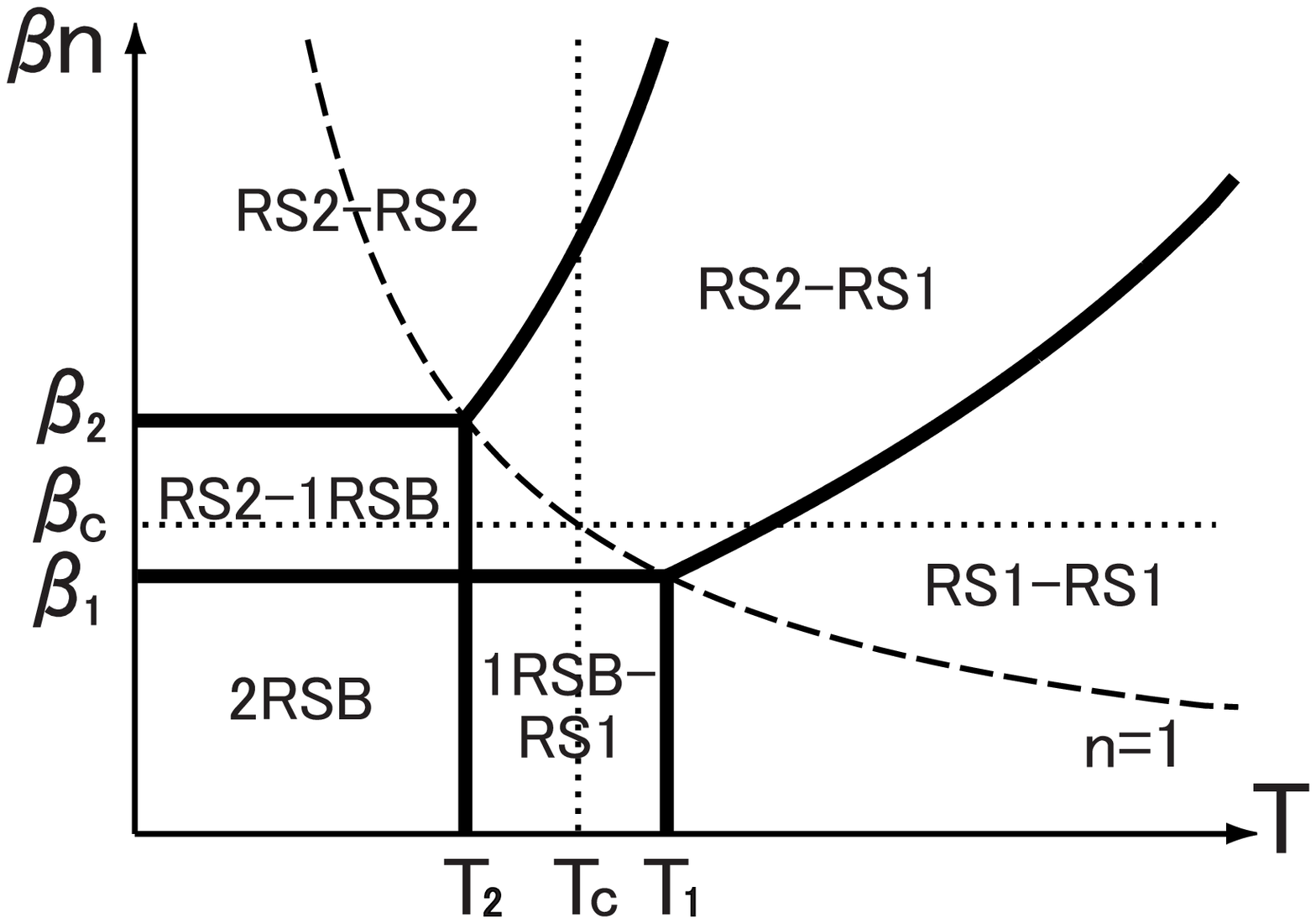}
\caption{Typical phase diagrams for $K=1$ (left) and $K=2$ 
 with $\ln\alpha_1/a_1<\ln\alpha_2/a_2$ and $\beta_2<\infty$
 (right).
 $T_{{\rm c},1,2}$ are defined as the inverse of 
 $\beta_{{\rm c},1,2}$ respectively.}
\label{fig:phase}
\end{figure}

 Next, we choose the correct solutions from the above 
 seven candidates of $\phi(\beta,n)$, 
 which depend on the values of parameters.
 We first summarize the case $K=1$ which is 
 naturally included in the above result.
 For $K=1$, the discrimination between the first and 
 the second hierarchies is useless, which means that 
 the correct solutions are chosen from the RS1-RS1, 
 RS2-RS2 and 1RSB-1RSB solutions 
 (hence abbreviated as RS1, RS2 and 1RSB in the $K=1$ case).
 When the solution of (\ref{eq:transitionK}) exists, 
 i.e.\ $R < \ln 2$ holds, 
 we have three phases on the $T$-$\beta n$ plane
 as investigated in~\cite{OgK1}. 
 The phase diagram in this case 
 is given in figure~\ref{fig:phase} (left). 
 On the other hand, for the case $\ln 2 \leq R$, 
 there is no phase transition and the RS1 solution 
 dominates the whole $T$-$\beta n$ plane, where there is no interest.

 In the case of $K=2$, the interesting case is again $R < \ln 2$, 
 i.e. $\beta_{\rm c}$ has a finite value. 
 Moreover, we should distinguish three cases depending on 
 the values of $\beta_{{\rm c},1,2}$.

 First, for $R_2 \le R_1$,
 the GDREM shows the same behavior as the standard discrete 
 REM, as discussed in~\cite{OTT}. 
 Hence, further investigation is not necessary in this case.
 
 Second, for the case $R_1 < \ln 2 < R_2 $, 
 where $\beta_2$ does not have a finite value,  
 we have three phases: the RS2-RS1, RS1-RS1 and 1RSB-RS1 phases. 
 In this case the second hierarchy is always in the RS1 phase and 
 only the first hierarchy shows phase transitions.
 In other words the system exhibits a similar phase structure as
 $K=1$. We focus on the properties of the hierarchical system here,
 and therefore we skip this case.

 The last and the most interesting case is 
 $R_1 <R_2<\ln 2$.
 This condition means that all the critical temperatures have finite values 
 and  $\beta_1<\beta_{\rm c}<\beta_2$. 
 In this case, there are six phases. 
 The resultant phase diagram is depicted 
 in figure~\ref{fig:phase} (right).
 To obtain this phase diagram, 
 we basically determine the contributing phase
 by the maximization principle based on the saddle-point method.
 In addition, we need some mathematical and physical criteria
 such as the continuity of $\phi(\beta,n)$ and 
 the non-negativity of entropy~\cite{OTT}.
 For instance, let us return to $K=1$ for simplicity.
 The boundary curve between the RS1 and RS2 phases is
 obtained by equating $\phi(\beta,n)$ for both the phases.
 The vertical phase boundary between the RS1 and 1RSB phases
 is derived by considering entropy crisis, which
 is identified with spin-glass transition as widely known.
 The horizontal boundary between the RS2 and 1RSB phases 
 should also exist as described in~\cite{OgK1}: 
 the RS2 phase cannot reach the quenched limit $\beta n=0$ 
 because it leads to unphysical behavior,
 e.g. $\lim_{n\to 0}\phi(\beta,n)/n \to \infty$.
 Thus, the dominant phase should naturally shift to other phase
 in decreasing $\beta n$. These discussions can also be
 applied to $K=2$, where the RSB of multiple step occurs and consequently
 partial entropy crisis is observed as mentioned in~\cite{OTT}.

 In the subsequent sections, we move on to the discussions of 
 lossy data compression and channel coding. 
 Actual evaluation of the performance of the random code ensemble 
 is conducted in the range $R_1 < R_2 < \ln 2$. 
 The phase diagram (figure~\ref{fig:phase}) and 
 the function $\phi(\beta,n)$ are of great use for this analysis,
 which explicitly demonstrates the advantage of the replica method.

\section{Lossy data compression}
\label{sec:datacomp}
 We start with the review of lossy data compression in~\cite{Merhav}.
 This issue has also been investigated by statistical
 mechanics~\cite{Murayama,HosakaKabashimaNishimori,MurayamaOkada,
 HosakaKabashima},
 and we concentrate on the hierarchical code here.
 After establishing how the generating function $\phi(\beta,n)$
 relates to lossy data compression, we apply the result
 of the replica analysis in the previous section.

\subsection{Distortion rate}
\label{sec:distortion}

 We prepare the hierarchical random code ensemble $\cal C$ with
 size $2^N$ and $aN$-bit hierarchical sequences $(a>1)$
 as in~\cite{Merhav}
 or equivalently in section~\ref{sec:bitrep}.
 For an arbitrary $aN$-bit hierarchical sequence,
 we represent it by one of the elements in $\cal C$,
 which amount to the process of lossy data compression. 
 $2^N$ sequences out of $2^{aN}$ possible ones have 
 one-to-one correspondence with one of the elements
 in $\cal C$, whereas others are distorted.
 To assess the performance of the compression process,
 we define the distortion (exactly the Hamming distortion)
 for the signal $\bm x$ as
\be
 \label{eq:distortion}
 \Delta(\bm x) \equiv \min_{\hat{\bm x} \in C}
 \left( d_{\rm H}(\bm x, \hat{\bm x}) - \frac{aN}{2} \right),
\ee
 where $\bm x$ and $\hat{\bm x}$ are $aN$-bit sequences.
 Subtraction of $aN/2$ in the definition is 
 for simplification of the analysis. 
 For extracting more information with regard to the distortion,
 it is an appropriate manner to define a characteristic function
 for the distortion~\cite{Merhav},
\be
 \Psi(s) = \left[\exp(-s \Delta({\bm x}))\right]_{\bm x, {\cal C}},
\ee
 that is, the moment generating function of the distortion.
 The brackets $[\ ]_{\bm x, {\cal C}}$ denote the average over 
 $\bm x $ and the ensemble of the code. 
 Actually, we may fix the bit sequence $\bm x$ as $\bm x=\bm 0$ 
 and remove average over $\bm x$, 
 because we take the average over the random code ensemble 
 $[\ ]_{{\cal C}}$. In the large $aN$ limit, 
 the rate of $\Psi(s)$, denoted by $\psi(s)$ and defined as follows,
 characterizes the performance of the random code ensemble,
\be
 \psi(s) \equiv - \lim_{N \rightarrow \infty} \frac{\ln \Psi(s)}{aN}
 = - \lim_{N \rightarrow \infty}
 \frac{\ln \left[\exp(-s \Delta(\bm 0))\right]_{\cal C}}{aN}.
\ee
 This distortion rate $\psi(s)$ has a direct relation with the generating 
 function $\phi(\beta,n)$ of the GDREM. 
 To see this, we should remember that the partition function 
 of the GDREM, $Z(\beta)$, can be written in the bit representation. 
 The distortion $\Delta(0)$ then corresponds to the ground state energy 
 of the GDREM. 
 Accordingly, the following transformation leads to the relation with 
 the replicated partition function of the GDREM:
\be
 \exp(-s \Delta(\bm 0))
 &=& 
 \lim_{n \rightarrow 0} \left(
 \sum_{\bm x \in {\cal C}}
 \exp \left\{ - \frac{s}{n}
 \left(
 d_{\rm H} (\bm x, \bm 0) 
 - \frac{aN}{2} \right) \right\} \right)^{n} \no \\
 &=&
 \lim_{n \rightarrow 0} Z^{n} \left( \frac{s}{n} \right).
\ee
 After taking average over the hierarchical random code ensemble,
 we have
\be
\label{eq:distortionrate}
 \psi(s)= - \lim_{N \rightarrow \infty} \lim_{n \rightarrow 0} 
 \frac{1}{aN} \ln \left[ Z^{n} \left(\frac{s}{n} \right)
 \right]_{{\cal C}} =  - \lim_{n \rightarrow 0} \frac{1}{a}
 \phi \left( \frac{s}{n},n \right).
\ee
 Consequently, we can directly assess the distortion rate $\psi(s)$ from 
 the generating function $\phi(\beta,n)$ in the replica analysis.
 
 To summarize, the distortion rate is accessible from 
 the replica analysis using the function $\phi(\beta,n)$ 
 with the constraint $s = \beta n$ and the limit of $n \rightarrow 0$.
 This means that the contributing phases to $\psi(s)$ are on the
 $\beta n$ axis in the $T$-$\beta n$ diagram,
 where there exist phase transitions 
 with respect to $s=\beta n$ as we see in section \ref{sec:replica}. 
 As a result, those transitions lead to the changes of the functional form 
 of the distortion rate. 

\subsection{Result}

 In the case of lossy data compression, the parameter $R=\ln 2/a$,
 which controls the phase transitions of the GDREM, 
 has the significance as the compression rate. 
 Since we deal with compression of data, the compression rate
 should be smaller than $\ln 2$, in which case the RSB transitions occur 
 as shown in section \ref{sec:replica}.

\subsubsection{$K=1$}

 To calculate the distortion rate $\psi(s)$, we take the limit 
 $n \rightarrow 0$ with keeping $\beta n=s$ in dealing with 
 the function $\phi(\beta,n)$. Accordingly, contributing phases 
 in the current problem turn out to be the RS2 and 1RSB phases.
 Using (\ref{eq:phiK2}) and (\ref{eq:distortionrate}),
 the distortion rate can be derived as
\be
 \psi(s) =  \left\{
 \ba{lll}
 - \frac{s}{2} \tanh \frac{s_R}{2}
 & ({\rm 1RSB}) & {\rm for} \ 0 \le s < s_R   \\
 - \ln \cosh \frac{s}{2} - R
 & ({\rm RS2})  & {\rm for} \ s_R \le s, 
 \ea\right.
\ee
 where the transition point $s_R$ is given from (\ref{eq:transitionK}),
\be
\label{eq:sR}
 R + \ln \cosh \frac{s_R}{2} 
 - \frac{s_R}{2} \tanh \frac{s_R}{2}
 = 0.
\ee
 The above solution coincides with the result in~\cite{Merhav}.
 Summarizing, the transition of the
 distortion rate is interpreted as the phase transition 
 on the $\beta n$ axis in the $T$-$\beta n$ diagram, namely
 the transition between the RS2 and 1RSB phases.

\subsubsection{$K=2$}
 We consider the case $R_1<R_2<\ln 2$ as mentioned 
 in section \ref{sec:replica}.
 As in figure~\ref{fig:phase}, there exist three phases
 on the $\beta n$ axis, the RS2-RS2, RS2-1RSB and 2RSB. 
 Substituting these solutions into (\ref{eq:distortionrate}), we have
\be
 \psi(s) =  \left\{
 \ba{ll}
 \frac{s}{2} \left( - \frac{a_1}{a} \tanh \frac{s_{R_1}}{2} 
 - \frac{a_2}{a} \tanh \frac{s_{R_2}}{2}
 \right) & {\rm (2RSB)} \\ 
 & {\rm for} \ 0 \le s < s_{R_1} \\ 
 - \frac{a_1}{a} \ln \cosh \frac{s}{2}
 -\frac{a_1}{a}R_1
 -\frac{s}{2} \frac{a_2}{a}  \tanh \frac{s_{R_2}}{2}
 & {\rm (RS2-1RSB)} \\
 & {\rm for} \ s_{R_1} \le s < s_{R_2} \\
 -\ln \cosh \frac{s}{2} - R
 & {\rm (RS2-RS2)} \\
 & {\rm for} \ s_{R_2} \le s, 
 \ea\right.
\ee
 where $s_{R_1}$ and $s_{R_2}$ are the solutions of equation (\ref{eq:sR})
 with substitutions $R=R_1$ and $R=R_2$, respectively.
 This also coincides with the result in~\cite{Merhav}. 

\subsection{Discussion}
 To judge whether the hierarchical structure reinforces
 the performance of the code in lossy data compression or not, 
 we compare the averaged distortion  
 $[\Delta({\bm 0})]_{\cal C}=\partial\psi/\partial s|_{s=0}$
 for both the cases $K=1$ and $2$. 
 The optimal case is the $K=1$ case, because it gives the smallest distortion.  
 This means that the introduction of the hierarchy of the current sort 
 has no positive effect on the lossy data compression, 
 which is the same conclusion as~\cite{Merhav}. 
 However, we here stress two advantages of our formulation. 
 
 First, our evaluation scheme is quite simple.
 We can treat both the cases $R_1 < R_2$ and $R_2 \leq R_1$
 in a unified framework and can easily see the relation
 between the cases $K=1$ and $K=2$. Generalization to the larger
 $K$ cases is also straightforward, whereas such a generalization
 seems to involve many technical difficulties in the original analysis. 
 
\begin{figure}[htb]
\begin{center}
\includegraphics[width=0.50\columnwidth]{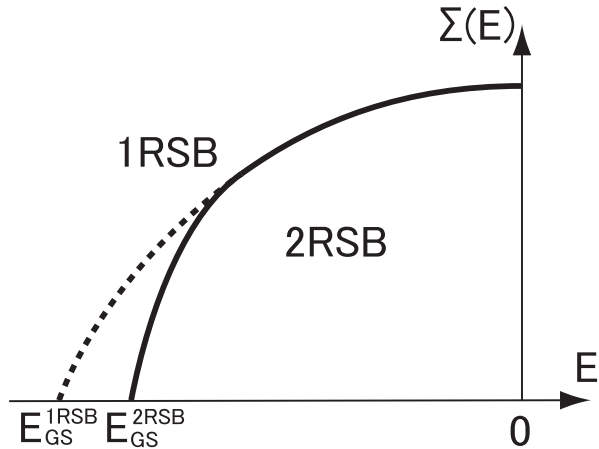}
\caption{A schematic behavior of complexity as a function of energy.
 Complexities from the 1RSB and 2RSB solutions are drawn by
 the dashed and the solid curves, respectively.
 $E^{\rm 1RSB}_{\rm GS}$ and $E^{\rm 2RSB}_{\rm GS}$ are the ground
 state energies of the 1RSB and 2RSB phases, respectively.
 For high energy states both the solutions give the same value of
 complexity, whereas for low energy states the 2RSB solution yields
 the smaller one.}
\label{fig:sigma}
\end{center}
\end{figure}

 Second, in our approach the transitions observed in the distortion
 rate can be understood as phase transitions with respect
 to the replica number, which include the RSB. This can provide more
 useful insights to signal processing including lossy data compression.
 For example, we can apply the complexity analysis to the current
 problem. The complexity, denoted by $\Sigma(E)$
 in figure \ref{fig:sigma}, is defined as the logarithm of 
 the number of pure states (see~\cite{OTT,Monasson} for details),
 which has a similar meaning to the entropy. Generally speaking,
 the higher step RSB leads to a decrease in low energy states,
 which implies the rise in ground-state energy (figure \ref{fig:sigma}). 

 This directly elucidates the performance loss of the hierarchical
 random code ensemble, because the distortion $\Delta(0)$ is
 identical with the ground-state
 energy of the GDREM. This observation implies that  
 the higher step RSB generally degrades the performance 
 in lossy data compression.  

\section{Channel coding}
\label{sec:channelcoding}
 In this section we move on to the problem of channel coding and
 see the relation with the replica method.
 Although the basic line of the analysis here is the same
 as in~\cite{OgK1}, there is a difference in the discussion
 of bound for the indicator function.

\subsection{General framework}

 Consider the BSC with reverse probability $p$ $(0 \le p \le 1/2)$.
 Following the framework in section~\ref{sec:bitrep},
 we also prepare an $N$-bit signal
 and encode it to a hierarchical $aN$-bit signal ($a>1$),
 which is included in the codebook $\cal C$ of size $2^{N}$.
 Then, we transmit an $aN$-bit code sequence 
 $\bm x$ through the BSC.
 The receiver decodes the original signal from an $aN$-bit output
 by the maximum likelihood decoding, which
 yields the inferred sequence $\bm y$.
 In the above setting, we define the error probability $P_{\rm E} (\cal C)$
 for a given set of original signal and codebook. 
 In particular, we focus on the value averaged over the set of codebook 
 $\cal C$, $\lsb P_{\rm E}({\cal C}) \rsb_{\cal C}$, whose expression is 
 given by
\be
\label{eq:avePe}
 \lsb P_{\rm E}({\cal C}) \rsb_{\cal C}
 = \sum_{\bm y} \lsb
 \sum_{\bm x \in {\cal C}}
 P(\bm x|{\cal C}) P(\bm y|\bm x)
 \Delta_{\rm ML}(\bm x, \bm y| {\cal C})
 \rsb_{\cal C}.
\ee
 $P(\bm x|\cal C)$ is the prior probability of
 the original signal $\bm x$ and $P(\bm y| \bm x)$
 is the posterior probability characterizing the BSC.
 $\Delta_{\rm ML}(\bm x,\bm y|\cal C)$ is the indicator 
 function of the maximum likelihood decoding, which is
 zero for successful decoding and unity for failure.

 For simplicity, we assume that the probability of the transmit
 signal $\bm x$ is uniform, $P(\bm x | {\cal C}) = 2^{-N}$.
 The posterior of the BSC is readily calculated as
\be
\label{eq:BSC}
 P(\bm y| \bm x) &=& (1-p)^{aN-{d_{\rm H}(\bm x, \bm y)}}
 p^{d_{\rm H}(\bm x, \bm y)} \no\\
 &=& \left( \frac{1}{2 \cosh (F/2)} \right)^{aN}
 \exp\left\{ -F \lb d_{\rm H}(\bm x, \bm y) - \frac{aN}{2}
 \rb \right\},
 \ee
 where $F=\ln \{(1-p)/p\}$
 (the Nishimori condition~\cite{Nishimoribook,Nishimori}).
 Besides, we can take the summation over $\bm y$ in (\ref{eq:avePe})
 and replace the reference codeword $\bm y$ with $\bm 0$,
 because the factor in $\lsb~\rsb_{C}$ becomes independent of
 $\bm y$ due to the summation $\sum_{{\bm x}\in {\cal C}}$ 
 and average $\lsb~\rsb_{C}$. 
 Substituting these, we obtain
\be
 \left[P_{\rm E} ({\cal C})\right]_{\cal C}
 = && \frac{1}{2^N}\lb \frac{1}{\cosh (F/2)} \rb^{aN}
 \no \\ && \times 
 \left[ \sum_{\bm x \in {\cal C}}  
 \exp\left\{ -F \lb d_{\rm H}(\bm x, \bm 0) - \frac{aN}{2} \rb \right\}
 \Delta_{\rm ML} (\bm x, \bm 0| {\cal C}) \right]_{\cal C}.
\label{eq:avePe2}
\ee
 It is a formidable task to evaluate the indicator function directly,
 and its bound is usually discussed by using inequalities. 
 For the hierarchical random code ensemble, it is convenient
 to use some different inequalities for different $K$.

\subsection{Analysis and result}
 In the case of channel coding, the parameter $R=\ln 2/a$
 corresponds to the transmission rate. 
 For successful communication, we have $a>1$ or equivalently $R < \ln 2$, 
 where the RSB phases play significant roles in the GDREM like
 lossy data compression. 

\subsubsection{$K=1$}
 This case corresponds to the conventional random code, and 
 we make use of the inequality in Gallager's original 
 work~\cite{Gallager1965,Gallagerbook},
\be
\label{eq:Gallager}
 \Delta_{\rm ML} (\bm x, \bm y| {\cal C})
 \le \left\{
 \sum_{\tilde{\bm x} \in {\cal C} \backslash \bm x} 
 \left( \frac{P(\bm y | \tilde{\bm x})}{P (\bm y| \bm x)}
 \right)^{\lambda}
 \right\}^{n}.
\ee
 The symbol ${\cal C} \backslash \bm x$ means the codebook ${\cal C}$ 
 with $\bm x$ removed.
 In this inequality we can take the arbitrary non-negative real values
 of $\lambda$ and $n$,
 which should be optimized for the tightest upper bound.
 In Gallager's work $\lambda$ was fixed as $\lambda=1/(n+1)$
 by using Jensen's and H\"{o}lder's inequalities. In contrast, in our approach
 we can readily deal with $\lambda$ without fixing, which is one of 
 the advantageous points of the current analysis.
 Moreover, the condition $\lambda=1/(n+1)$ may lead to a looser bound 
 in some cases as indicated in~\cite{AJKMM,KabashimaSaad2004,Hayashi}. 
 Hence, we adopt the two-parameter optimization here.
 
 Insertion of (\ref{eq:BSC}) and (\ref{eq:Gallager}) into (\ref{eq:avePe2}) 
 yields
\be
\label{eq:genbound}
 \left[P_{\rm E} ({\cal C})\right]_{\cal C}
 &\le& \frac{1}{2^N}
 \left( \frac{1}{\cosh(F/2)} \right)^{aN} \no \\
 & & \times
 \left[ \sum_{\bm x \in {\cal C}}
 \exp\left\{-F(1-n\lambda)
 \left( d_{\rm H}(\bm x,\bm 0) - \frac{aN}{2} \right)
 \right\} \right. \no \\
 & & \times \left.
 \left( \sum_{\tilde{\bm x} \in {\cal C} \backslash \bm x} 
 \exp \left\{ -F(1-n\lambda)
 \left( d_{\rm H}(\tilde{\bm x},\bm 0) - \frac{aN}{2} \right)
 \right\}
 \right)^{n} \right]_{\cal C}.
\ee
 In the current case $K=1$, codewords are not mutually correlated, which
 allows us to rewrite the upper bound as
\be
\label{eq:separate}
 \left[P_{\rm E} ({\cal C})\right]_{\cal C}
 &\le& \frac{1}{2^N}
 \left( \frac{1}{ \cosh(F/2)} \right)^{aN} \no \\
 & & \times
 \left[ \sum_{ \bm x \in {\cal C}}
 \exp\left\{-F(1 - \lambda n)
 \left( d_{\rm H}(\bm x,\bm 0) - \frac{aN}{2} \right)
 \right\} \right]_{\cal C} \no \\
 & & \times \left[
 \left( \sum_{\tilde{\bm x} \in {\cal C} \backslash \bm x} 
 \exp \left\{ -F \lambda
 \left( d_{\rm H}(\tilde{\bm x},\bm 0) - \frac{aN}{2} \right)
 \right\}
 \right)^{n} \right]_{\cal C} \no \\
 &= &
 \frac{1}{2^N}
 \left( \frac{1}{ \cosh(F/2)} \right)^{aN}
 \exp ( N \lbb \phi(F(1-n\lambda),1) 
 +\phi(F \lambda,n) \rbb ). \no \\
\ee
 With regard to the last factor in the first line, the absence of $\bm x$  
 in the sum over codebook $\cal C$ can be neglected without
 loss of generality in the limit $N \rightarrow \infty$.
 Substituting the trivial expression of $\phi(F(1-n\lambda),1)$
 and optimizing $\lambda$ and $n$, we obtain the bound
\be
 \left[P_{\rm E} ({\cal C})\right]_{\cal C}
 &\le& 
 \exp (-aN E_{r}^{K=1}(R)),
\ee
 for $N \rightarrow \infty$, where $E_r^{K=1}(R)$
 is known as Gallager's error exponent for the BSC,
\be
\label{eq:errorexponent}
 E_r^{K=1} (R) = \mathop{\rm \max}_{0\le n, \lambda}
 \left\{-\frac{1}{a} \phi \left(F\lambda, n\right)
 - \ln \cosh \frac{F(1-\lambda n)}{2} + 
 \ln \cosh \frac{F}{2} \right\}.
\ee
 Next, we apply the RS1, RS2 and 1RSB solutions in (\ref{eq:phiK2})
 to the computation of the error exponent (\ref{eq:errorexponent}).

\begin{itemize}
\item{RS1}

 The error exponent is expressed as
\be
 E_r^{K=1}(R) & = & 
 \mathop{\large \rm max}_{0 \le n, \lambda} \left\{ - n R
 - n \ln \cosh \frac{F\lambda}{2}
 \right. \no \\
 & & \ \ \ \ \ \ \ \left. - \ln \cosh \frac{F(1-\lambda n)}{2}
 + \ln \cosh \frac{F}{2} \right\}.
\ee
 From the maximization condition with respect to $n$ we have 
\be
 R + \ln \cosh \frac{F\lambda }{2}
 - \frac{F\lambda}{2} \tanh \frac{F\lambda}{2} =0,
\ee
 leading $F\lambda =\beta_{\rm c}$.  
 Clearly, this means that the optimal solution is given on 
 the RS1-1RSB boundary if $0 \le n \le 1$. 
 Hence, we need not to take this solution into account
 because it can be included in the 1RSB solution. 
 For $n>1$ and $F\lambda =\beta_{\rm c}$, 
 the correct $\phi(F\lambda,n)$ is not given by the RS1, 
 which also leads to the irrelevance of the RS1 solution.

\item{RS2}

 In this case the error exponent is
\be
 E_r^{K=1}(R) & = &
 \mathop{\large \rm max}_{0 \le n, \lambda} \left\{ - R
 - \ln \cosh \frac{F\lambda n}{2}
 \right. \no \\
 & & \ \ \ \ \ \ \ \left. -\ln \cosh \frac{F(1-\lambda n)}{2}
 + \ln \cosh \frac{F}{2} \right\}. 
\ee
 This form allows us to consider the maximization 
 with respect to the product $\lambda n$, 
 giving $\lambda n=1/2$. The substitution yields 
\be
 E_r^{K=1}(R)
 =-R-2\ln\cosh \frac{F}{4}+\ln\cosh\frac{F}{2}.
\label{eq:errorRS2}
\ee
 Due to the functional form of the RS2 solution, where
 $\lambda$ and $n$ appear only in the product $\lambda n$, 
 we have a redundancy in optimizing $\lambda$ and $n$. 
 On the $T$-$\beta n$ plane, this redundancy means that all the points 
 on a line $\beta n =F/2$ in the RS2 phase give 
 the identical result (\ref{eq:errorRS2}). 
 As $F$ decreases, this horizontal line on the plane
 goes down along the vertical axis $\beta n$ 
 and finally this reaches the phase boundary between 
 the RS2 and 1RSB phases,
 which gives the bound of the RS2 solution $2\beta_{\rm c}<F$.
 This can be regarded as a simple graphical interpretation 
 of the behavior of the error exponent.

\item{1RSB}

 The error exponent is
\be
 E_r^{K=1}(R) &=& 
 \mathop{\large \rm max}_{0 \le n, \lambda} 
 \left\{ - \frac{F \lambda n}{2}
 \tanh \frac{\beta_{\rm c}}{2} - \ln \cosh \frac{F(1-\lambda n)}{2}
 \right. \no \\
 && \left. \hspace{4cm}
 + \ln \cosh \frac{F}{2} \right\}.  
\ee
 Maximization with respect to $\lambda n$ yields
 $1- \lambda n=\beta_{\rm c} / F$. 
 Substituting this, we have
\be
 E_r^{K=1}(R) &=& 
 - \frac{F}{2} \left( 1 - \frac{\beta_{\rm c}}{F} \right)
 \tanh \frac{\beta_{\rm c}}{2} - \ln \cosh \frac{\beta_{\rm c}}{2}
 + \ln \cosh \frac{F}{2}  \no \\
 &=& -\frac{F}{2} \tanh \frac{\beta_{\rm c}}{2} 
 + R + \ln \cosh \frac{F}{2}.
\ee
 Again, on the $T$-$\beta n$ plane, 
 this result is irrespective of temperature similar to the RS2.
 This solution gives $E_r=0$ at $F=\beta_{\rm c}$, 
 which is the bound of successful error correction
 for the infinite size limit.
\end{itemize}

 Summarizing the above results, we finally obtain
\be
\label{eq:exponent-DREM}
 E_r^{K=1}(R) = \left\{
 \ba{lll}
 -R - 2 \ln \cosh \frac{F}{4} & & \\
 \qquad +\ln \cosh \frac{F}{2}
 & {\rm (RS2)} & {\rm for}\ 2\beta_{\rm c}<F \\
 -\frac{F-\beta_{\rm c}}{2} \tanh \frac{\beta_{\rm c}}{2} 
 -\ln \cosh \frac{\beta_{\rm c}}{2}
 & & \\
 \qquad +\ln\cosh \frac{F}{2}
 & {\rm (1RSB)} 
 & {\rm for} \ \beta_{\rm c} \le F \le 2 \beta_{\rm c} \\
 0 & \hspace{-1.4cm} ({\rm 1RSB}, \beta n=0) & {\rm for} \ F<\beta_{\rm c}. 
 \ea\right. \no\\
\ee
 After some calculations, we can confirm that the above error exponent 
 by the replica analysis is in perfect agreement 
 with Gallager's expression~\cite{Gallager1965,Gallagerbook}, 
 as well as the consistency with Shannon's channel 
 coding theorem~\cite{Shannon1948a,Shannon1948b}. 

 In fact, the above result is not a novel one, 
 because the error exponent of some models in information theory, 
 including the random code ensemble,
 has already been investigated by using statistical mechanics 
 in some works~\cite{KabashimaSaad2004,KSNS,Montanari2001,SMSK,MoraRivoire}. 
 However, the methods used there are different from the one we proposed. 
 We here emphasize some advantages of our method.

 The first one is the applicability to larger $K$ cases, which will 
 be demonstrated in the next subsection. In such cases, the codewords are
 mutually correlated and the analysis becomes more complicated. Despite this, 
 our scheme can evaluate the error exponent without any approximation 
 except for a slight modification of Gallager's original inequality 
 (\ref{eq:Gallager}). 

 Second, as observed and will be observed again for larger $K$, 
 the transition of the error 
 exponent is the RSB between the RS2 and 1RSB phases,
 which provides a simple interpretation 
 of the function form change of the error exponent.
 This fact has never been observed or discussed explicitly, 
 which might be due to 
 the condition $\lambda=1/(n+1)$ originated from Jensen's and H\"{o}lder's 
 inequalities.   
 Moreover, our analysis reveals that the RSB transition with respect to 
 the replica number, which cannot be observed 
 from a thermodynamical quantity after the quenched average,
 is significant in channel coding. 
 This situation is similar to lossy data compression.
 
 In addition, from the above discussion, we can find that the RS1 phase is 
 excluded from contributing phases, which has a significance
 in successful decoding. 
 The detail will be argued after the analysis of $K=2$.

\subsubsection{$K=2$}
\label{sec:K=2}
 In this case, the codewords are mutually correlated, which invalidates 
 the factorization (\ref{eq:separate}). 
 In such a case, the general form (\ref{eq:genbound}) should be directly 
 evaluated, as long as Gallager's inequality (\ref{eq:Gallager}) is used. 
 This calculation can actually be done by a novel replica approach,
 which is somewhat different from the standard one. 
 In this approach, we deal with the factor in $[~]_{\cal C}$ 
 in (\ref{eq:genbound}) as the partition function of the GDREM
 with $n+1$ replicas. 
 However, there are two noteworthy points: 
 First, the inverse temperatures are not common to all 
 replicas. One replica out of $n+1$ has the inverse temperature
 $F(1-n \lambda)$, and others have $F\lambda$. This requires
 the special treatment of one replica with different temperature
 in the replica analysis. 
 Second, the correlation between the special one replica and other $n$ 
 replicas exists in this case. 
 As seen in the summation in (\ref{eq:genbound}),
 the special one cannot take the same state as those of other $n$ replicas. 
 Hence, we need to introduce asymmetry between replicas, 
 in addition to the RSB among non-special $n$ replicas. 
 This novel replica method will generally be applicable to any other codes 
 with mutual correlation among codewords, 
 under the situation that Gallager's inequality is used.
 
 Actually, we applied this novel method to $K=2$
 and obtained the bound of 
 the error probability. However this approach requires rather involved 
 calculations. Fortunately we can avoid this novel approach by 
 a slight modification of 
 Gallager's original inequality. In this paper,
 we demonstrate this simpler approach in the framework of 
 the ordinary replica analysis of the GDREM for $K=2$.
 We confirmed that the results from both the schemes
 coincide with each other.

 As stated, we use another inequality for the indicator function here, 
 instead of Gallager's original inequality (\ref{eq:Gallager}), as
\be
\label{eq:Gallager2}
 & &
 \Delta_{\rm ML} (\{\bm x_1, \bm x_2(\bm x_1)\},
 \bm y | {\cal C})
 \no \\
 &\le&
 \ \ \ \left(
 \sum_{\{ \bm x_1, \tilde{\bm x}_2 (\bm x_1) \}
 \in {\cal C} \backslash \{\bm x_1, \bm x_2 (\bm x_1)\}} 
 \left( \frac{P( \bm y |
 \{ \bm x_1, \tilde{\bm x}_2 (\bm x_1) \}) }
 {P (\bm y | \{\bm x_1, \bm x_2(\bm x_1)\})}
 \right)^{\lambda_1}
 \right)^{n_1}
 \no \\
 & & +\left(
  \sum_{\{\tilde{\bm x}_1, \tilde{\bm x}_2 (\tilde{\bm x}_1) \}
 \in {\cal C} \backslash \{\bm x_1, \tilde{\bm x}_2 (\bm x_1)\}} 
 \left( \frac{P(\bm y |
 \{ \tilde{\bm x}_1, \tilde{\bm x}_2 (\tilde{\bm x}_1) \}) }
 {P (\bm y| \{\bm x_1, \bm {x_2}(\bm x_1)\})}
 \right)^{\lambda_2}
 \right)^{n_2},
\ee
 where the codeword $\bm x$ is represented 
 by the hierarchical components as 
 $\bm x = \{ \bm x_1, \bm x_2 (\bm x_1)\}$, 
 and $\{ \bm x_1, \tilde{\bm x}_2 (\bm x_1) \}$ is a codeword whose
 first block is the same as the correct transmission codeword $\bm x$.
 Note that the second-block codeword depends on the first-block
 one, which is denoted by $\bm x_2 (\bm x_1)$.
 
 Substituting (\ref{eq:Gallager2}) into (\ref{eq:avePe2}), 
 we can assess the upper bound of the error probability. 
 The contribution from the first term of (\ref{eq:Gallager2}) is
 equal to
\be
 &&
 \frac{1}{\alpha_{2}^{N}} \left( \frac{1}{ \cosh (F/2)} \right)^{a_2 N}
 \no \\
 && \times
 \left[ \sum_{\bm x_2 \in {\cal C}_2}
 \exp \left\{ - F  (1- \lambda_1 n_1)
 \left( d_{H}( \bm x_2 , \bm 0 ) - \frac{a_2N}{2} \right) \right\}
 \right]_{{\cal C}_2} \no \\
 && 
 \times \left[ \ \ \ \left( \sum_{\tilde{\bm x}_2
 \in {\cal C}_2 \backslash  \bm x_2 } 
 \exp \left\{ -F \lambda_1 \left( d_{H}(\tilde{\bm x_2},\bm 0)
 - \frac{a_2N}{2} \right) \right\} \right)^{n_1} \right]_{{\cal C}_2},
\label{eq:first_cont}
\ee 
 where we express the entire codebook ${\cal C}$ by the
 hierarchical codebooks as ${\cal C}=({\cal C}_1,{\cal C}_2)$,
 then perform the summations and ensemble average
 with respect to the first hierarchy,
 $\sum_{{\bm x}_{1}}$ and $[~]_{{\cal C}_1}$. 
 The sizes of codebooks ${\cal C}_1$ and ${\cal C}_2$ are $\alpha_1^N$ 
 and $\alpha_2^N$, respectively. 
 The statistical independence between two different codebooks
 in the first hierarchy is essential for deriving
 expression (\ref{eq:first_cont}). 
 The absence of correlation between ${\bm x}_2$ 
 and $\tilde{{\bm x}}_2 \neq {\bm x}_2$ is necessary as well.
 From the result of $K=1$,
 this contribution (\ref{eq:first_cont}) is simply expressed as 
 $\exp (- a_2 N E_{r}^{K=1} (R_2))$.

 The contribution from the second term is
\be
 &&
 \frac{1}{2^N} \left( \frac{1}{ \cosh (F/2)} \right)^{aN}
 \no \\
 && \times
 \left[ \sum_{(\bm x_1, \bm x_2(\bm x_1) ) \in {\cal C}}
 \exp \left\{ -F(1-\lambda_2 n_2)
 \left( d_{H} (\{ \bm x_1, \bm x_2 (\bm x_1) \}, \bm 0 )
 -\frac{aN}{2} \right) \right\}
 \right]_{\cal C}
 \no \\
 && \times
  \left[ \left(
 \sum_{\{ \tilde{\bm x}_1, \tilde{\bm x}_2 ( \tilde{\bm x}_1) \}
 \atop \in {\cal C} \backslash \{\bm x_1, \tilde{\bm x}_2 (\bm x_1)\}} 
 \exp \left\{ -F \lambda_2
 \left(
 d_{H}( \{ \tilde{\bm x}_1, \tilde{\bm x}_2 (\tilde{\bm x}_1) \}, \bm 0)  
 - \frac{aN}{2} \right) \right\} \right)^{n_2} \right]_{\cal C}.
 \no \\
\ee
 To derive this expression, the absence of correlation between
 $\bm x_1$ and $\tilde{\bm x}_1\neq {\bm x}_1$ is used in a similar manner.
 As a result, this contribution is represented by 
 the generating function of the GDREM for $K=2$. Denoting this contribution by
 $\exp ( -a N E_{r}^{K=2} (R,R_1,R_2))$, we can write the exponent as
\be
 \label{eq:secondterm}
 E_{r}^{K=2} (R,R_1,R_2) &=&
 \mathop{\rm \max}_{0\le n, \lambda}
 \left\{  - \frac{1}{a} \phi \left( F \lambda , n \right)
 - \ln \cosh \frac{F(1-\lambda n)}{2} \right. \no \\
 && \left. \hspace{4cm} 
 + \ln \cosh \frac{F}{2} \right\}.
\ee
 Note that $\phi(F\lambda,n)$ in (\ref{eq:secondterm}) is
 for the $K=2$ case.
 Hence, the bound of the error probability for $K=2$
 is expressed as
\be
 \label{eq:finalresult}
 [P_E(E)]_{\cal C} 
 \le \exp (- a_2 N E_{r}^{K=1} (R_2)) + \exp (- a N E_{r}^{K=2}
 (R,R_1,R_2)),
\ee
 for $N \rightarrow \infty$.
 Therefore, we must compare the two contributions by computing 
 $E_r^{K=1} (R_2)$ and $E_r^{K=2} (R,R_1,R_2)$.
 
 Here we give a comment on (\ref{eq:finalresult}). 
 For $K=2$, an intuitive discussion in~\cite{Merhav} suggests 
 the following result
\be
 \label{eq:Merhavresult}
 [P_E(E)]_{\cal C} 
 \le \exp (- a_2 N E_{r}^{K=1} (R_2)) + \exp (- a N E_{r}^{K=1} (R)),
\ee
 for $N \rightarrow \infty$, which differs from (\ref{eq:finalresult}).
 However, as mentioned later, the difference between 
 (\ref{eq:finalresult}) and (\ref{eq:Merhavresult}) is irrelevant 
 because both the bounds are identical for $N \rightarrow \infty$.
 Despite this irrelevance, 
 we consider that
 our result is more natural and suggestive. The reason is as follows:
 In the case of the hierarchical random code ensemble,
 we can expect that the error probability
 can be decomposed into failure event from respective hierarchy.
 Our result (\ref{eq:finalresult}) based on (\ref{eq:Gallager2})
 clearly reflects this feature. 
 In the same way, we can expect the bound of the error probability 
 for general $K$ as
\be 
 [P_E(E)]_{\cal C} \leq
 \sum_{\nu=1}^{K}\exp\lb -\sum_{j=0}^{\nu -1}(a_{K-j} )N E_{r}^{K=\nu} \rb,
\ee
 for $N \rightarrow \infty$. 
 This expression should be confirmed in future works.

 Next we compute the bound (\ref{eq:finalresult}) using the result of
 the replica analysis. 
 The first term has already been estimated in the analysis of $K=1$, 
 and here we evaluate the second term.
 As stated in section \ref{sec:replica}, 
 the case of $R_1 (< R) < R_2 < \ln 2$ is dealt with here. 
 The detail of the analysis is in~\ref{sec:errorexponentK2}. 
 We summarize the main result in the following.

 Contributing phases to the error exponent are 
 the RS2-RS2, RS2-1RSB and 2RSB. 
 The error exponent is obtained as 
\be
\label{eq:exponentK=2}
 & & E_r^{K=2}(R,R_1,R_2) \no \\
 &=& \left\{
 \ba{ll}
 -R-2\ln \cosh \frac{F}{4} + \ln \cosh \frac{F}{2} 
 & {\rm(RS2-RS2)} \\
 & {\rm for}\ 2\beta_2<F \\
 -\ln \cosh \frac{1}{2}(F- \beta_x) 
 - \frac{a_1}{a} R_1 & \\
 \quad - \frac{a_1}{a} \ln \cosh \frac{\beta_x}{2} 
 - \frac{a_2}{a} \frac{\beta_x}{2} \tanh \frac{\beta_2}{2} & \\
 \quad + \ln \cosh \frac{F}{2} 
 & {\rm (RS2-1RSB)} \\
 & {\rm for}\  \beta_1 + \beta_y \le F < 2 \beta_2 \\
 -\frac{F-\beta_y}{2} \tanh \frac{\beta_y}{2}
 -\ln\cosh \frac{\beta_y}{2} & \\
 \quad + \ln \cosh \frac{F}{2} & {\rm (2RSB)} \\
 & {\rm for}\ \beta_y \le F \le \beta_1 + \beta_y
 \\ 0 & ({\rm 2RSB}, \beta n=0)  \\
 & {\rm for}\ F \le \beta_y,
 \ea \right.
\ee
 where $\beta_x$ and $\beta_y$ are given by
 \be
 & & a_1 \tanh \frac{\beta_x}{2} + a_2 \tanh \frac{\beta_2}{2}
 = a \tanh \frac{F-\beta_x}{2}, \no \\
 & & a_1 \tanh \frac{\beta_1}{2} + a_2 \tanh \frac{\beta_2}{2}
 = a \tanh \frac{\beta_y}{2}.
\ee
 As a result, only the phases on the $\beta n$ axis contribute
 to the error exponent similar to lossy data compression.
 This property will hold for arbitrary $K$, 
 which simplifies the analysis for larger $K$.

 Finally we must compare the contributions 
 from the first and the second terms on right hand side of
 (\ref{eq:finalresult}). We checked it numerically and
 concluded that the first term always dominates when
 $R_1 <R_2 <\ln 2$. On the other hand, for $R_2 \leq R_1$,
 the second term dominates and yields the same result as $K=1$.
 
\subsection{Discussion}

 Now we are ready to compare the performances of
 $K=1$ (non-hierarchical) and $K=2$ for $R_1 < R_2 < \ln 2$.
 Comparing the dominant contribution $\exp(-a_2 N E_r^{K=1} (R_2))$
 for $K=2$ with the non-hierarchical result $\exp(-a N E_r^{K=1} (R))$,
 we found that the error exponent of the non-hierarchical code
 always surpasses that of the hierarchical code for fixed $N$.
 Therefore, the hierarchy degrades the performance of decoding
 for $R_1 \le R_2$. Although this conclusion is the same as
 Merhav's discussion~\cite{Merhav}, our formulation has
 crucial advantages in the way of reasoning. 

 Our approach has never loosened the bound of the error exponent,
 except for Gallager's inequality (\ref{eq:Gallager}).
 This can be achieved with the aid of the replica method,
 and furthermore the two-parameter optimization
 with respect to $\lambda$ and $n$ can be reasonably conducted.
 In conventional approaches, several inequalities,
 such as H\"older and Jensen inequalities, are employed as
 in~\cite{Merhav}.
 However, in such analyses the parameter optimization is usually performed
 only with respect to $n$, by fixing $\lambda$ as $\lambda=1/(n+1)$.
 These manipulations do not only loosen the bound of the error exponent
 but also obscure the origin of transitions of the error exponent.
 Without such risks, our formulation enables us to analyze
 the performance of random codes in detail.
 As a result, some physical significances of the behavior of
 the error exponent can be extracted as follows.
 
 We know that the contributing phases to the error exponent
 always include the RS2 phase and/or the 1RSB phase 
 in their hierarchy, and the RS1 phase is excluded.
 The reason is probably elucidated as follows. 
 In a successful case of decoding, the concentration of the Gibbs measure
 to a certain input signal is expected to be realized.
 In the RS2/1RSB phases such concentration actually occurs:
 for the RS2 a single state is chosen by definition
 (see figure \ref{fig:saddlepoint} or~\cite{OTT,OgK1})
 and for the 1RSB the measure concentration occurs due to glassy nature.
 On the other hand, for the RS1 phase such concentration
 does not occur or the phase is paramagnetic,
 which corresponds to an inefficient decoding.
 Consequently we do not need to consider the RS1 phase
 for the discussion of optimal decoding.
 We expect this argument is applicable to a general code ensemble.

 The above observation also gives some benefits
 in the practical analyses of $\phi(\beta,n)$:
 this function will always be written by the function
 of the product $\beta n$ for the contributing RS2/1RSB phases.
 This is due to the measure concentration for successful decoding,
 in which case the replicated partition function is written
 with the product $\beta n$. From this discussion,
 we also conclude that the condition $\lambda = 1 / (n+1)$
 for the one-parameter optimization, which is used in the original
 discussion~\cite{Gallager1965,Gallagerbook} and valid there,
 sometimes yields a looser upper bound than the two-parameter optimization.
 Actually, if we put $\lambda=1/(n+1)$ in (\ref{eq:finalresult}),
 the RS1 phase becomes included in the final solution,
 and we obtain a looser bound than (\ref{eq:exponentK=2}),
 although it does not contribute to the error probability
 for $N \rightarrow \infty$ due to the dominance of
 the first term $\exp(-a_2 N E_r^{K=1} (R_2))$ in the current situation.
 Hence, the one-parameter optimization used in Gallager's work
 should be carefully examined. This observation will also be helpful to 
 other problems of channel coding.

\section{Conclusion}

 We investigated the hierarchical random code ensemble
 by using the direct relation with the GDREM. 
 We sketched how the replica analysis is carried out and 
 is useful for large deviation analysis, namely
 computations of the distortion rate in lossy data
 compression and Gallager's error exponent in channel coding.
 We provided formulae for these quantities
 and demonstrated how they are evaluated in the case of
 two hierarchy levels. 
 For lossy data compression, the distortion rate from 
 the replica analysis is in perfect agreement with~\cite{Merhav}.
 Using our method, we could calculate the distortion rate
 quite readily, which is one of the advantageous points.
 We also interpreted the behavior of the distortion rate
 in terms of the complexity, and found that the emergence of
 the higher step RSB degrades the performance of data compression.
 From the relation between the complexity and the RSB transition,
 this conclusion will hold for a general hierarchical code,
 which is helpful in designing code.
 In addition, 
 we obtained the novel result for channel coding. 
 The procedure to compute the upper bound 
 is different from Gallager's argument.
 This difference arises from the correlation between
 codewords in the replica analysis. 
 Our result from the two-parameter optimization seems quite natural
 because the measure concentration is associated with optimal
 performance as we discussed.

 In both the problems, we argued that the RSB transition 
 with respect to the replica number is significant.
 Although our analysis is based on the mapping between
 two fundamental models in signal processing and spin glasses,
 we expect the application of the proposed method 
 to other ensembles or problems in signal processing.
 We also hope that the observation in this paper for data compression
 or channel coding from the viewpoint of the RSB will be of use
 in other problems. 
 Such an application will be our future work.

\section*{Acknowledgments}

 The authors are grateful to Y Kabashima for useful discussions. 
 TO is supported by a Grant-in-Aid
 Scientific Research on Priority Areas `Novel
 State of Matter Induced by Frustration' (19052006 and 19052008).
 K Takeda is supported by
 a Grant-in-aid Scientific Research on Priority Areas
 `Deepening and Expansion of Statistical Mechanical Informatics
 (DEX-SMI)' from MEXT, Japan no 18079006. 

\appendix
\section{Error exponent for $K=2$}
\label{sec:errorexponentK2}
 We evaluate the error exponent in each phase for the case $K=2$ by
 using (\ref{eq:secondterm}) and the result of the replica analysis
 (\ref{eq:phiK2}). 

\begin{itemize}
\item{RS1-RS1}

 This gives the same result as the RS1 of $K=1$.

\item{RS2-RS1}

 From (\ref{eq:phiK2}),
\be
 && E_r^{K=2} (R,R_1,R_2) \no \\
 &=& \mathop{\rm max}_{0 \le n,\lambda} \left\{
 - \left(
 R_1 + \frac{a_1}{a} \ln \cosh \frac{F n \lambda}{2} \right)
 - n \left(
 R_2 + \frac{a_2}{a} \ln \cosh \frac{F \lambda}{2} \right)
 \right. \no \\
 & & \left.
 - \ln \cosh \frac{F(1-n\lambda) }{2} + \ln \cosh \frac{F}{2}
 \right\}.
\ee
 Two extremization conditions give
 $ F \lambda =  \beta_2$, which means that
 the extremization region is always on the boundary
 between the RS2-RS1 and RS2-1RSB phases, and
\be
\label{eq:extremumRS2RS1}
 \frac{a_1}{a} \tanh \frac{n \beta_2}{2}
 + \frac{a_2}{a}
 \tanh \frac{\beta_2}{2}
 = \tanh \frac{F - n \beta_2}{2}.
\ee
 The error exponent is rewritten as
\be
 && E_r^{K=2} (R,R_1,R_2) \no \\
 &=& - \left(
 R_1 + \frac{a_1}{a} \ln \cosh \frac{\beta_2 \hat{n}}{2} \right)
 - \frac{a_2}{a} \frac{\beta_2 \hat{n}}{2} \tanh \frac{\beta_2}{2} 
 \no\\
 & & - \ln \cosh \frac{F 
 - \beta_2 \hat{n} }{2} + \ln \cosh \frac{F}{2},
\ee
 where $\hat{n}$ is the solution of (\ref{eq:extremumRS2RS1}).
 As mentioned, the extremization region is on the phase boundary,
 and this result can be included in the case of the RS2-1RSB.

\item{RS2-RS2}

 This gives the same result as the RS2 for $K=1$.

\item{1RSB-RS1}

 From (\ref{eq:phiK2}),
\be
 && E_r^{K=2} (R,R_1,R_2) \no \\
 &=&
 \mathop{\rm max}_{0 \le n,\lambda} \left\{
 -\frac{a_1}{a} \frac{F n \lambda}{2} \tanh \frac{\beta_1}{2}
 - n \left( R_2 + \frac{a_2}{a}
 \ln \cosh \frac{F \lambda }{2} \right) \right.
 \no \\
 & & \left.
 - \ln \cosh \frac{F(1- n \lambda)}{2} + \ln \cosh \frac{F}{2}
 \right\}.
\ee
 From two extremization conditions we have $F\lambda=\beta_2$, 
 which means that the extremization region is always on the boundary 
 between the 1RSB-RS1 and 2RSB phases, and
\be
\label{eq:extremum1RSBRS1}
 \frac{a_1}{a} \tanh \frac{\beta_1}{2}
 + \frac{a_2}{a}
 \tanh \frac{\beta_2}{2}
 = \tanh \frac{F - n \beta_2}{2}.
\ee
 Then the error exponent is changed to
\be
 && E_r^{K=2} (R,R_1,R_2) \no \\
 &=& - \frac{\beta_2 \hat{n}}{2}
 \left( \frac{a_1}{a} \tanh \frac{\beta_1}{2}
 + \frac{a_2}{a} \tanh \frac{\beta_2}{2} \right) \no\\
 & & - \ln \cosh \frac{F - \beta_2 \hat{n} }{2} 
 +\ln \cosh \frac{F}{2},
\ee
 where $\hat{n}$ is the solution of (\ref{eq:extremum1RSBRS1}).
 As stated, the extremization region is on the phase boundary.
 This result can be included in the case of the 2RSB.

\item{RS2-1RSB}

 From (\ref{eq:phiK2}),
\be
 && E_r^{K=2} (R,R_1,R_2) \no \\
 &=& \mathop{\rm max}_{0 \le n,\lambda} \left\{
 - \frac{a_1}{a} R_1 - \frac{a_1}{a} \ln \cosh \frac{F \lambda n}{2}
 - \frac{a_2}{a} \frac{F \lambda n}{2} \tanh \frac{\beta_2}{2}
 \right.
 \no \\
 & & \left.
 - \ln \cosh \frac{F (1- \lambda n)}{2} + \ln \cosh\frac{F}{2}
 \right\}.
\ee
 The extremization condition with respect to $\lambda n$ gives
\be
\label{eq:extremumRS21RSB}
 \frac{a_1}{a} \tanh\frac{F \lambda n}{2}
 + \frac{a_2}{a} \tanh \frac{\beta_2}{2}
 = \tanh\frac{F(1-\lambda n)}{2}.
\ee
 Substituting the solution of (\ref{eq:extremumRS21RSB}),
 we have a part of solution (\ref{eq:exponentK=2}).

\item{2RSB}

 From (\ref{eq:phiK2}),
\be
 && E_r^{K=2} (R,R_1,R_2) \no \\
 &=&
 \mathop{\rm max}_{0 \le n,\lambda} \left\{
 -\frac{F \lambda n}{2} \left( \frac{a_1}{a} \tanh \frac{\beta_1}{2}
 + \frac{a_2}{a} \tanh \frac{\beta_2}{2} \right) \right.\no\\
 & & \left.
 - \ln \cosh \frac{F (1- \lambda n)}{2} + \ln \cosh \frac{F}{2}
 \right\}.
\ee
 From the extremization with respect to $\lambda n$,
\be
\label{eq:extremum2RSB}
 \frac{a_1}{a} \tanh \frac{\beta_1}{2}
 + \frac{a_2}{a} \tanh \frac{\beta_2}{2}
 = \tanh \frac{F(1-\lambda n)}{2}.
\ee
 Inserting the solution of (\ref{eq:extremum2RSB}),
 we have a part of solution (\ref{eq:exponentK=2}).

\end{itemize}


\end{document}